\documentclass[pra,twocolumn,showpacs,amsfonts,amsmath,floatfix]{revtex4}

\usepackage[pdftex]{graphicx}
\usepackage{array}
\usepackage{ulem}

\begin{document}
\title{A Test Resonator for Kagome Hollow-Core Photonic Crystal Fibers for Resonant Rotation Sensing}

\author{Ihsan Fsaifes$^{1}$}
\author{Gilles Feugnet$^{2}$ }
\author{Alexia~Ravaille$^{1,2,3}$}
\author{Beno\^it Debord$^{4}$ }
\author{Fr\'ed\'eric G\'er\^ome$^{4}$ }
\author{Assaad Baz$^{4}$ }
\author{Georges Humbert$^{4}$ }
\author{Fetah Benabid$^{4}$ }
\author{Sylvain Schwartz$^{2}$ }
\author{Fabien Bretenaker$^{1}$}
\email{Fabien.Bretenaker@u-psud.fr}
 \affiliation {$^1$Laboratoire Aim\'e Cotton, Universit\'e Paris-Sud, ENS Cachan, CNRS, Universit\'e Paris-Saclay, Orsay, France
 \\$^2$Thales Research \& Technology, Palaiseau, France\\
 $^3$Thales Avionics, Ch\^atellerault, France\\
 $^4$GPPMM group, Xlim research institute, CNRS-Universit\'e de Limoges, Limoges, France}


\markboth{Photonics Technology Letters,~Vol.~xx, No.~xx, xxx}%
{Shell \MakeLowercase{\textit{et al.}}: Bare Demo of IEEEtran.cls for Journals}


\begin{abstract}
We build ring resonators to assess the potentialities of Kagome Hollow-Core Photonic Crystal Fibers for future applications to resonant rotation sensing. The large mode  diameter of Kagome fibers permits to reduce the free space fiber-to-fiber coupling losses, leading to cavities with finesses of about 30 for a diameter equal to 15 cm. Resonance linewidths of  3.2~MHz with contrasts as large as 89\% are obtained. Comparison with  7-cell photonic band gap (PBG) fiber leads to better finesse and contrast with Kagome fiber. Resonators based on such fibers are compatible with the angular random walk required for medium to high performance rotation sensing. The small amount of light propagating in silica should also permit to further reduce the Kerr-induced non-reciprocity  by at least three orders of magnitudes in 7-cell Kagome fiber compared with 7-cell PBG fiber.
\end{abstract}
\maketitle



\section{Introduction}

Recent progresses in hollow-core photonic crystal fibers (HC-PCF) and compact single frequency laser sources renewed interest for the resonant fiber optic gyroscope (R-FOG). The basic principle of an R-FOG is to measure, by using an external probe laser, the eigenfrequencies of two counter-propagating modes of a fiber ring cavity, their difference being proportional to the angular velocity of the device (Sagnac effect). Unlike the interferometric fiber optic gyroscope, which requires several hundreds of meters of fiber to reach the required sensitivity for inertial navigation, here, only a very short length (few tens of cm to few meters) of optical fiber is sufficient. While the first proof-of-principle of an R-FOG was made as early as 1983 \cite{Meyer1983}, its impact on applications remained limited owing to several reasons such as high level of Kerr induced bias instabilities in solid-core fibers.

The breakthrough of using bandgap guidance to trap light in a low-index core makes HC-PCFs promising candidates for building medium to high grade gyroscopes \cite{Terrel2012}. Hollow-core (HC) fibers, where light propagates mostly in air, have the advantage of significantly reducing the Kerr effect, known as a strong limitation on the bias stability of the R-FOG \cite{Kim2006}. These fibers exhibit reduced bulk scattering and detrimental effects (thermal fluctuations, mechanical stress and radiation effects) that create noise and dissipate signals \cite{Blin2007,Zhao2014}, although new problems emerge due to surface scattering \cite{Terrel2012, Wu2015}.
A first demonstration of a R-FOG using a 7-cell core hollow-core photonic band gap (PBG) fiber has been published by Terrel et al \cite{Terrel2012}. The achieved performances (bias stability around 1 deg/s) are still several orders of magnitude poorer than the one typical required for inertial navigation. Recently, Qiu et al. \cite{Qiu2014} have reported a R-FOG based on standard PM fiber with a laser stabilized to an external resonator. The bias stability and angular random walk were 0.1 deg/hr and 0.008 deg/$\sqrt{\mathrm{hr}}$, respectively. These interesting results are a promising sign that commercial R-FOG with navigation or tactical grade performance might be attainable in the near future. 

The HC fibers investigated here are based on a Kagome lattice design. Unlike PBG guiding fibers, photons are confined inside Kagome HC fibers not by PBG but via a mechanism akin to Von Neumann-Wigner bound states in a continuum, and whereby the fiber guided modes cohabitate with those of the cladding without notably interacting. This inhibited coupling between the propagation mode and the surrounding silica cladding is explained by the large transverse spatial-phase mismatch between the core and the cladding fields. The mode power fraction propagating in glass is reported to be below 0.05\% \cite{Couny2007,Humbert2004}, which is about two orders of magnitude lower compared to typical values (few percents) for PBG fibers. As a result of this weak interaction between the core and cladding modes, Kagome HC-PCFs exhibit broad transmission regions with relatively low loss covering the spectral range from IR to UV, and open new potentialities for many applications \cite{Wang2013,Debord2014}, in particular for rotation sensing with a weak Kerr effect, and potential reduction of backscattering \cite{Terrel2012}.

In this paper, we present experimental results on an Inhibited-Coupling (IC) guiding Kagome HC-PCF based resonators. In particular, by using a simple semi-bulk cavity design, we build test resonators that allow us to characterize and investigate different HC fibers and to compare their performances in terms of cavity linewidth and fraction of intensity in silica, for gyroscope application. Two figures of merit are defined to evaluate them.

%

\section{Fundamental limits in R-FOGs}\label{sec02}
Ultimately, if we assume that all effects related to laser noise, backscattering, mechanical and thermal drifts, etc. are avoided, the noise performance of the R-FOG is limited by the shot-noise limit (SNL). The latter can be made in principle arbitrarily small by increasing the intra-cavity power, but this will in return increase non-linear effects within the fiber, such as Kerr effect, leading to bias instalility. Consequently, the maximum achievable signal-to-noise ratio will be a trade-off between increasing the circulating power and minimizing the nonlinear effects. In this section, we express the driving parameters to obtain this trade-off for the case of hollow-core fibers

\subsection{Shot-noise sensitivity limit}
The frequency difference $\Delta f$ between the two counter-propagating modes of the R-FOG cavity is given by:
\begin{equation}
\Delta f=\frac{4 A}{\lambda L}\dot{\theta}\ ,\label{eq01}
\end{equation}
where $A$ is the area enclosed by the cavity and $L$ its optical perimeter, while $\lambda$ and $\dot{\theta}$ are the light wavelength and the rotation rate, respectively. We note here that the scale factor is directly proportional to the ratio $A/L$, and thus cannot be increased by multiplying the number of fiber turns in the cavity, contrary to what happens in interferometric fiber optic gyros \cite{Schwartz2014}.  Thus, for simplicity, we replace $A/L$ by $D/4$, where $D$ is the diameter of the equivalent circular cavity. For a rotation measurement involving $N_{\mathrm{Ph}}$ photons, the minimum measurable rotation rate $\dot{\theta}_{\mathrm{min}}$ is:  
\begin{equation}
\dot{\theta}_{\mathrm{min}}=\frac{\lambda L}{4 A}\delta f_{\mathrm{min}}=\frac{\lambda}{D}\frac{\Gamma}{\sqrt{\alpha N_{\mathrm{Ph}}}} ,\label{eq02}
\end{equation}
where $\delta f_{\mathrm{min}}$ is the minimum measurable value of the frequency difference $\Delta f$, $\Gamma$ and $\alpha$ are respectively the cavity resonance linewidth and contrast. This contrast, which is between 0 and 1, is defined as $(I_{\mathrm{max}}-I_{\mathrm{min}})/I_{\mathrm{max}}$,  where $I_{\mathrm{max}}$ and $I_{\mathrm{min}}$ are respectively the maximum and minimum intensities for the considered transmission or reflection resonance. The number of photons used for the measurement can be related to the detected optical power $P_{\mathrm{det}}$ and the measurement time $T$ by $N_{\mathrm{Ph}} = P_{\mathrm{det}} T / (hc/\lambda)$, so that the expression of the angular random walk (ARW) is \cite{Schwartz2014}:
\begin{equation}
ARW=\dot{\theta}_{\mathrm{min}}\sqrt{T}=\frac{\Gamma}{D}\sqrt{\frac{hc\lambda}{\alpha P_{\mathrm{det}}}}\ .\label{eq03}
\end{equation}
This equation shows that the ARW can be minimized by increasing the contrast and decreasing the linewidth of the cavity. This leads to define a cavity figure of merit $FOM_{\mathrm{SNL}}$ for the shot noise limit:
\begin{equation}
FOM_{\mathrm{SNL}}=\frac{\sqrt{\alpha}}{\Gamma}\ .\label{eq04}
\end{equation}
This figure of merit permits to compare the different fibers, even if a maximum contrast is not always optimal for sensitivity to rotation. It is also worth noticing that the finesse of the cavity is not the relevant figure of merit to discuss the sensitivity to rotation.
\subsection{Kerr effect}
For given figure of merit (eq. \ref{eq04}), cavity diameter $D$, and wavelength $\lambda$, the only way to improve the ultimate performance of an R-FOG is to increase the optical power launched inside the cavity. However, this increases the Kerr effect in the fiber, which turns out to be an important source of bias for the R-FOG, as discussed below.

In the presence of an intensity $I$, the refractive index $n$ of silica evolves according to  $n = n_0+n_2 I$, where $n_0$ is the linear refractive index and $n_2$ is the nonlinear refractive index coefficient. This results in two different nonlinear refractive indices if the two counter-propagating fields inside the fiber cavity have a power difference $\Delta P$, inducing the following bias for the optical gyroscope \cite{Chow1985}:
\begin{equation}
\Delta\dot{\theta}=\frac{c n_2}{Dn_0}\frac{\Delta P \eta}{\sigma}\ ,\label{eq05}
\end{equation}
where $\sigma$ is the effective area of the guided mode and $\eta$ the fraction of the mode power which is propagating in silica. Table \ref{table01} gives the maximum values of the relative power difference $\Delta P/P$ (for $P = 1\;\mathrm{mW}$ inside the cavity) to achieve a bias $\Delta\dot{\theta}<0.001\;^{\circ}/\mathrm{hr}$, a typical value for a high performance grade gyroscope, calculated for  three different fibers: standard telecom fiber, HC PBG fiber, and HC Kagome fiber. These calculations were performed for $D = 15\;\mathrm{cm}$ and $n_2=3.0\times10^{-20}\;\mathrm{m}^2/\mathrm{W}$ for silica. The value of $\eta$ for the Kagome fiber at $1.5\,\mu$m is taken from \cite{Couny2007}. The refractive index for SMF 28 fiber is taken equal to 1.5. The values used for MFDs and $\eta$ in HC PBG and HC Kagome fibers in table \ref{table01}  are just typical orders of magnitudes. The precise values of these parameters depend on the details of the design of the fibers, as will be described below (see table \ref{table02}) for the fibers we actually used in our experiments.

\begin{table}[htbp]
\centering
\caption{\bf Typical values of Kerr effect for different types of fibers}
\begin{tabular}{cccc}
\hline
Fiber & $\eta$ & MFD & $\Delta P/P$ for 0.001 $^{\circ}$/hr bias \\
\hline
SMF 28 & 1.0 & 10 $\mu$m & $\simeq 9\times10^{-6}$ \\
HC PBG & 0.05 & 10 $\mu$m & $\simeq 1\times10^{-4}$\\
HC Kagome & 0.0004 & 40 $\mu$m & $\simeq 0.2$ \\
\hline
\end{tabular}
  \label{table01}
\end{table}

HC PBG fibers lead to a significant reduction of Kerr effect with respect to standard SMF fiber. However, we can see that for the Kagome fiber, which presents the largest mode field diameter and the lowest mode power fraction inside silica, the bias of the R-FOG becomes almost insensitive to any power difference between the two counter propagating beams. Differently, when the mode field diameter is small like for a standard fiber or a seven-cell hollow core PBG fiber, the condition on the relative power difference between the two modes inside the cavity becomes mores stringent. Of course, if the intra-cavity power is larger than 1 mW, the relative power difference $\Delta P/P$ in Table~\ref{table01} is reduced accordingly. Thus there is a trade-off between increasing the optical power to reduce the shot noise limit and the acceptable relative power difference for the Kerr effect. From this analysis, we can define a second figure of merit for the Kerr effect induced bias: 
\begin{equation}
FOM_{\mathrm{Kerr}}=\frac{\sigma}{\eta}\ .\label{eq06}
\end{equation}

To summarize, the mode field diameter of the fiber should be relatively large to avoid bias caused by the Kerr effect. The principle of Kagome fibers, which relies on small values of $\eta$, is favorable to the gyrometer performance. 

\section{Resonator design}\label{sec03}
\begin{figure}[htbp]
\includegraphics[width=\columnwidth]{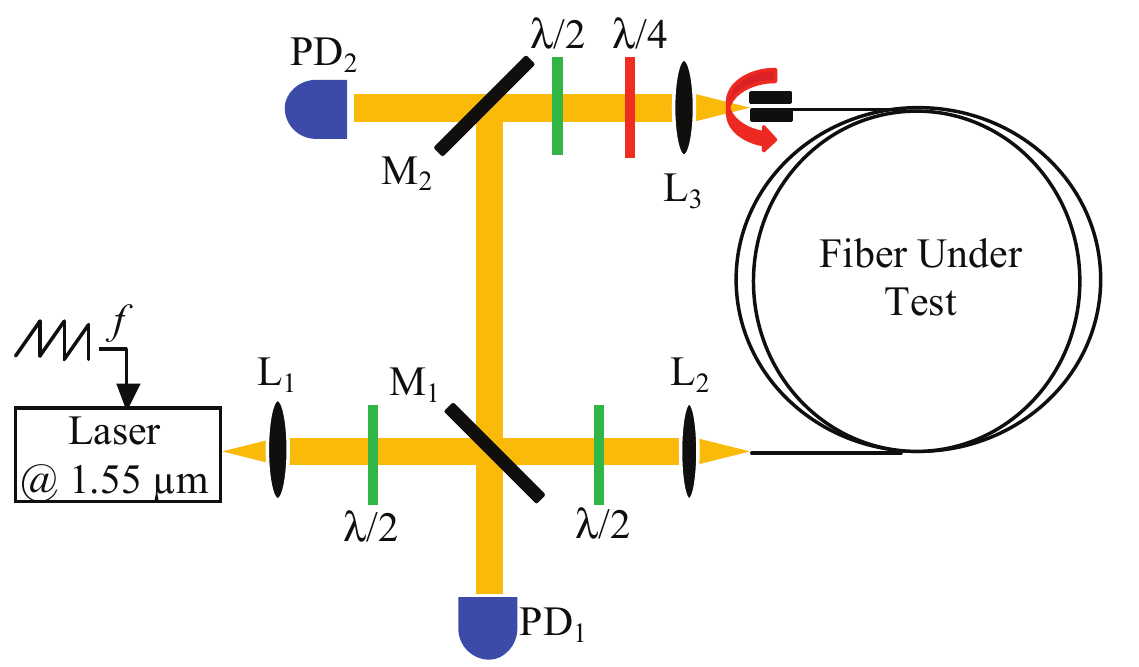}
\caption{Experimental setup. PD$_1$ and PD$_2$ are photodetectors,measuring the intensities reflected and transmitted by the cavity, respectively. L$_1$, L$_2$, and L$_3$ are lenses. M$_1$ and M$_2$ are the cavity mirrors. One end of the fiber is mounted on a high precision rotation mount, as shown by the arrow.}\label{Fig01}
\end{figure}

As stated above, HC fibers are a very promising solution to the problem of Kerr effect occurring in R-FOGs. However, all-HC-fiber couplers are not yet available and splicing a HC fiber to a standard fiber \cite{Terrel2012} induces splicing losses of the order of 1 dB per splice \cite{Aghaie2010,Hayes2016} and  backreflection. Like in reference \cite{Sanders2006}, we choose to build a test cavity based on free space couplers (see Fig.~\ref{Fig01}), with mirrors at 45$^{\circ}$ incidence acting as input and output couplers. Lenses L$_2$ and L$_3$ are used to couple light in and out of the fiber.

This semi-bulk design allows for an easy optimization of the transmission of mirror M$_1$ to get as close as possible to the so-called critical coupling (M$_1$ transmission equal to the cavity internal losses, including the transmission of M$_2$), which maximizes the contrast of the resonances. Mirror M$_2$ is highly reflecting. Its very small transmission is just used to monitor the intracavity power, when necessary. The free space section of the cavity also allows to insert birefringent elements to control the overall cavity polarization eigenstates, without using polarization maintaining fiber. One can then obtain a stable single polarization resonance. 

\section{Kagome HC-PCF design and properties}\label{sec04}
\begin{figure}[htbp]
\includegraphics[width=\columnwidth]{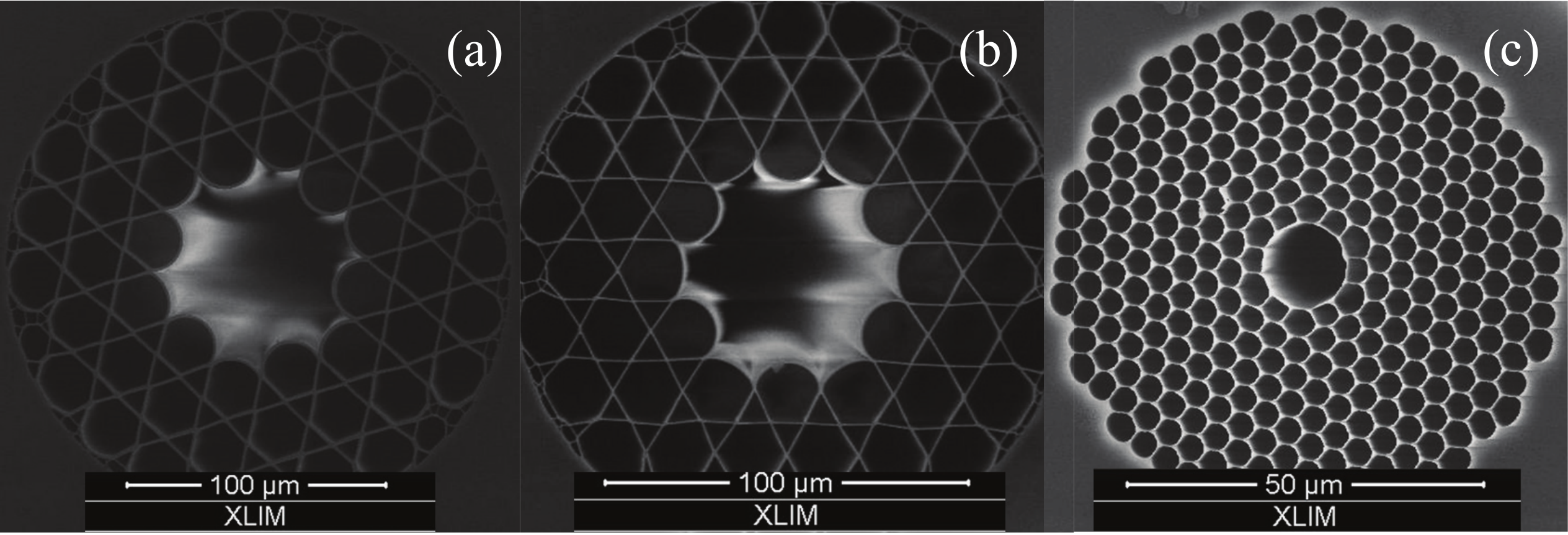}
\caption{SEM pictures of the (a) Kagome `A' fiber, (b) Kagome `B' fiber, and (c) 7-cell PBG fiber used here.}\label{Fig02}
\end{figure}
Two 7-cell HC Kagome fibers [see Figs.\,\ref{Fig02}(a,b)] have been designed, fabricated, and tested into the cavity, and compared to a 7-cell HC PBG fiber [see Fig.\,\ref{Fig02}(c)].  The two Kagome fibers have been selected to provide both low-loss guidance at 1550 nm and different core diameters. The fibers parameters, length, attenuation at 1.55 $\mu$m, mode field diameter at $1/e^2$ (MFD), fraction of mode power in silica $\eta$, and the overall coupling efficiency $P_{\mathrm{out}}/P_{\mathrm{in}}$ which takes into account coupling into the fiber and the propagation losses are summarized in Table \ref{table02}

The Kagome fibers labeled `A' and `B' exhibit hypocycloid contour with negative curvature parameters $b=0.8$ and $0.74$ respectively (see \cite{Debord2013} for the definition of $b$), and silica strut thicknesses of 1250~nm and 480~nm, respectively. The fractions $\eta$ and MFDs were calculated through the modal solver of a commercial software based on the finite-element method with an optimized anisotropic phase-matching layer (PML) \cite{Selleri2001} for the three fibers as displayed in Table  \ref{table02}. 

\begin{figure}[htbp]
\includegraphics[width=0.8\columnwidth]{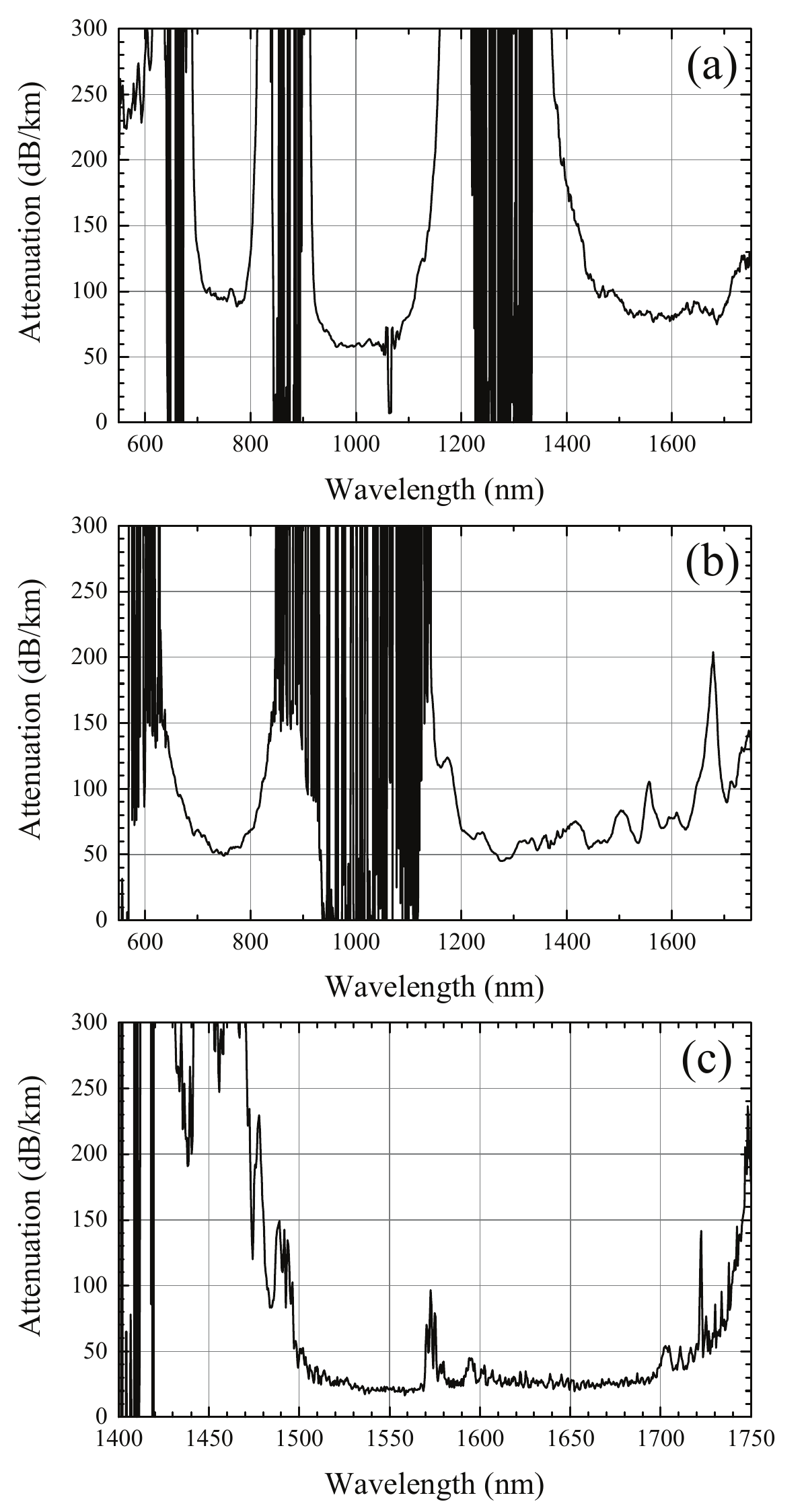}
\caption{Measured propagation losses spectra for the (a) Kagome `A' fiber, (b) Kagome `B' fiber, and (c) 7-cell PBG fiber used here. The measurements are performed using the cut-back method.}\label{Fig02bis}
\end{figure}

The loss spectra of the three hollow-core fibers used here are shown in Fig.\ \ref{Fig02bis}. Finally, the sensitivity to bend of these two fibers has been experimentally measured to be respectively equal to 0.3 dB/turn and to 0.1 dB/turn at 1.55 $\mu$m for a 15~cm curvature diameter.

\section{Experimental setup and results}\label{sec05}
We use a narrow band ($\simeq$10 kHz) fiber-coupled single-frequency polarized laser operating at 1.55 $\mu$m (see Fig. \ref{Fig01}). The laser frequency is linearly swept to probe the resonances of the fiber resonator. The resonator is formed by two free-space mirrors (M$_1$, M$_2$), of 90 \% and 98.5 \% reflectivity, respectively. The total cavity length is equal to the sum of the fibre length and the free space path. Coupling into the fiber is performed with a pair of antireflection coated aspheric lenses (L$_2$, L$_3$) allowing a good matching between the external laser mode and the fiber mode field. The tested fibers are wound on a 15-cm-diameter coil. Their end faces are cleaved with an angle to minimize back-reflections. Two photodetectors (PD$_1$, PD$_2$) are used to respectively monitor both the reflected and transmitted intensities \cite{Fsaifes2015}.

\begin{table}[htbp]
\centering
\caption{\bf Properties of the fibers tested here}
\begin{tabular}{cccccc}
\hline
Fiber & Length  & Losses & MFD & $\eta$  & $\dfrac{P_{\mathrm{out}}}{P_{\mathrm{in}}}$ \\
 &  (m) &(dB/km) & ($\mu$m) & (\%) & (\%) \\
 \hline
Kagome A & 2.4 & 80 & 40 & $5.0\times 10^{-2}$ & 91\\
Kagome B & 2.3 & 60 & 34 & $2.0\times 10^{-3}$ & 85\\
7-cell PBG & 19 & 22 & 7.7 & 6 & 70\\
\hline
\end{tabular}
  \label{table02}
\end{table}

\begin{figure}[htbp]
\includegraphics[width=0.75\columnwidth]{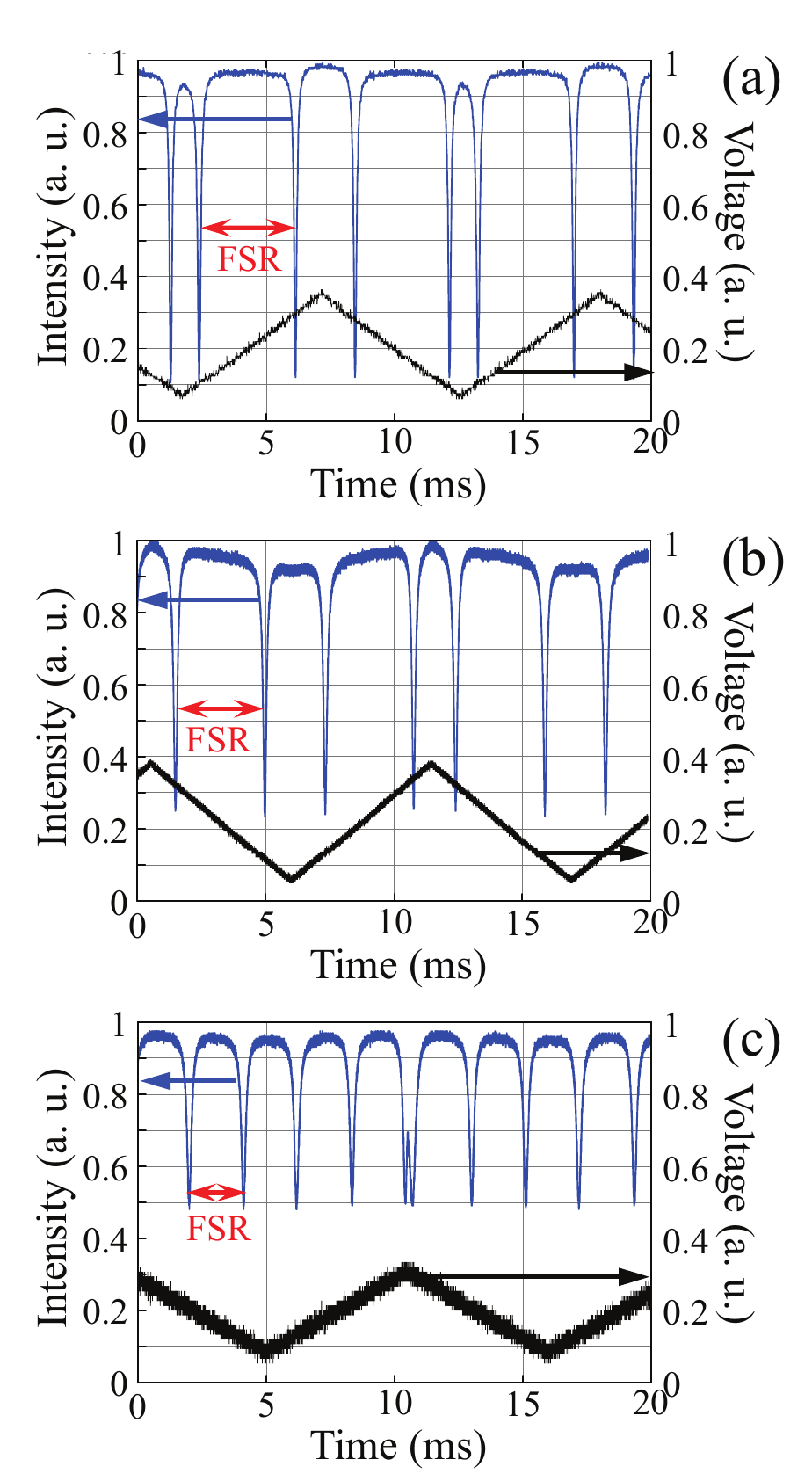}
\caption{Evolution of the intensity reflected by the cavity versus time for (a) Kagome A, (b) Kagome B, (c) 7-cell PBG fiber. The ramp applied to the laser frequency, at a frequency close to 90~Hz, is also displayed. The measured cavity FSRs and linewidths are respectively (a) 106 and 3.2 MHz, (b) 109 and 5.2 MHz, (c) 15.5 and 1.7~MHz.}\label{Fig03}
\end{figure}

 The cavity resonance peaks recorded in reflection on PD$_1$ are shown in Fig. \ref{Fig03}. The HC Kagome fiber A gives the best results in terms of contrast (89 \%) and finesse (33) compared to the Kagome B (78 \% and 21) and the 7-cell PBG (50 \% and 9) fibers. The Kagome A and B fibers have roughly the same lengths (cavity FSRs equal to 106 and 109 MHz, respectively), while in the case of the 7-cell PBG fiber, we had to take a much longer fiber (FSR = 15.5 MHz) in order to get a cavity linewidth in the MHz range. This is due to a lower coupling efficiency in this latter case. For the Kagome A fiber, the ratio between the input and output power through the fiber is ~91 \% so that after taking into account propagation losses (about 6 \%) and lenses losses (about 1 \%), the total losses that can be attributed to free-space to fiber coupling are about 2 \%. For the 7-cell PBG hollow core fiber, because of its small mode field diameter (MFD) of about 8~$\mu$m, the coupling efficiency is only about 70 \%, so by taking into account propagation losses (about 9 \%) and lenses losses (about 1 \%), the total losses attributed to free-space to fiber coupling are about 20~\%. Notice also that it is possible to obtain higher finesses with much shorter lengths of 7-cell PBG fibers, as shown in \cite{Sanders2006}, but at the cost of a much larger cavity linewidth.
 
The fibers that we use are not polarization maintaining. We thus measured the depolarization in our Kagome fiber coils by injecting a linear polarization and measuring the best extinction we can achieve at the output through a quarter-wave plate and a linear polarizer. This leads to a depolarization ratio of -16~dB, showing that the fiber can mainly be considered as a single mode birefringent element.  Thus, as mentioned in section \ref{sec02}, using quarter- and half-wave plates in the free space section of the cavity, associated with an optimized orientation of the fiber end, one obtains a single resonance peak for all the fibers. This demonstrates the possibility to select a single spatial and polarization mode of the cavity, despite the fact that the fibers were not specifically designed to be polarization maintaining. This single mode behavior is of outmost importance for gyroscope operation and was associated with a good stability of the finesse and contrast of the cavity, especially considering the fact that no special care was taken to isolate the cavity from mechanical vibrations and thermal fluctuations.
 
Table \ref{table03} gives, for each fiber, the calculated values for the two figures of merit previously defined.  For the sake of comparison, we have also added a line giving the figures for solid-core fiber, using the cavity parameters mentioned in \cite{Qiu2014}, and assuming that the cavity resonance contrast $\alpha$ is equal to 1.

\begin{table}[htbp]
\centering
\caption{\bf Figures of merit of the tested fibers, in comparison with solid-core single-mode fiber}
\begin{tabular}{ccc}
\hline
Fiber & Figure of merit  & Figure of merit  \\
 &  for shot noise &for Kerr effect  \\
 & (MHz$^{-1}$) & ($\mu$m$^2$) \\
 \hline
Kagome A & 0.29 & $2.5\times 10^6$ \\
Kagome B & 0.17 & $4.5\times 10^7$ \\
7-cell PBG & 0.42 & 780 \\
Solid-core  SMF & 7.5 & 79\\
\hline
\end{tabular}
  \label{table03}
\end{table}

The interesting result from these comparisons is that, even if the Kagome fibers presents higher propagation losses, the 7-cell PBG and Kagome fibers are relatively close in terms of shot noise limit (there is only a factor of two between the respective figures of merits). This is due to the better coupling efficiency demonstrated with the Kagome fibers that somehow compensates for the higher propagation losses. These results illustrate also the fact that the propagation losses are not the only parameter to take into account when choosing a fiber and that the coupling efficiency into the fiber and the fiber-to-fiber coupling efficiency are also primordial to get a high cavity performance. However, when comparing the figure of merit for the Kerr effect limit, the Kagome fibers seem more attractive than the PBG fiber. This would be even more so with Kagome fibers with lower loss figures such as the one reported in \cite{Debord2013}.

With the Kagome A fiber, assuming a cavity diameter of $D = 15$ cm, $\lambda = 1.55\;\mu$m, and $P_{\mathrm{det}} = 1$ mW, the shot noise limit is about $0.0014\;^{\circ}/\sqrt{\mathrm{hr}}$, which compares favorably with ref. \cite{Qiu2014} and is compatible with inertial navigation requirement. In such a fiber, Kerr effect induced bias become completely negligible.

Of course, in such a semi-bulk architecture based on hollow-core fiber, there is still a long way to go before reaching the shot noise limit in a rotation measurement. Indeed, back-scattering inside the fiber and mechanical drifts are serious problems that need to be closely analyzed. However, we still believe that shot noise, which is the ultimate limit to the gyro random walk, is worth considering.

\section{Conclusion}
We have experimentally investigated the use of a Kagome hollow-core photonic crystal fiber to build resonators. More precisely, we have analyzed the possibility to build such a resonator with a linewidth, a resonance contrast, and a fraction of light in silica compatible with future applications to resonant fiber optic gyroscopes.  

This work stresses the fact that, for an optimized design of an R-FOG, fibers with low propagation losses, high input coupling efficiency, and single spatial and polarization mode are required. Contrary to PBG HC fiber, Kagome HC fibers exhibit a relatively large core size, a very low spatial overlap of the core mode with the silica core-wall, and a coupling efficiency exceeding 90 \%. Despite the fact we used Kagome fibers with higher loss figures than that of PBG fiber, these attributes make hollow core fibers based on Kagome lattice design a good candidate for resonant fiber-optic gyroscopes, especially from the point of view of Kerr effect induced bias. However, further studies are required to investigate the role of backscattering \cite{Ma2011,Terrel2012,Wu2015} and modal purity in resonators based on Kagome fibers, and also to improve the cavity ruggedness by connectorizing or splicing HC fibers with limited losses, before one can perform real rotation measurements based on such fibers.

\section*{Acknowledgments}

This work is supported by the Agence Nationale de la Recherche (Project PHOBAG: ANR-13-BS03-0007 and PhotoSynth: ANR-11-CHEX-0009) and the European Space Agency (ESA).



\end{document}